# Light-Sculpted Azopolymer Colloids:
# From Patchy Spheres to Porcupine and Pineapple Morphologies

Sh. Golghasemi Sorkhabi [1,2,3,*], R. Barille [1], M. Loumaigne [1], A. Korbut [4], S. Zielinska [4], and E. Ortyl [4]

[1] University of Angers/UMR CNRS 6200, MOLTECH-Anjou, 49045 Angers, France

[2,*] Department of Physics, Bilkent University, 06800 Ankara, Turkey

[3,*] UNAM National Nanotechnology Research Center and Institute of Materials Science and Nanotechnology, Bilkent University, 06800 Ankara, Turkey

(email: Shahla.golghasemi@gmail.com)

[4] Wroclaw University of Technology, Faculty of Chemistry, Department of Polymer Engineering and Technology, 50-370 Wroclaw, Poland

**Abstract**
A simple optical strategy to transform patchy PMMA–azopolymer composite nanoparticles into complex, fully three-dimensional morphologies using controlled laser polarization is presented. The particles consist of a PMMA core decorated with nanoscale azopolymer patches that undergo localized photofluidization upon trans–cis isomerization. Linear polarization drives directed mass transport within each patch, producing elongated super-cones that collectively yield porcupine-like particles, whereas circular polarization generates isotropic bump deformations reminiscent of sea-pineapple structures. A nonlinear, volume-conserving geometric model quantitatively reproduces the patch-to-filament transition. Brownian and Jeffery-flow simulations reveal that these photoinduced morphologies dramatically alter hydrodynamic behavior, leading to enhanced anisotropic diffusion, reduced rotational randomization, and polarization-dependent transport amplification in shear flow. This light-driven, reversible sculpting method provides a versatile route to programmable colloidal shapes and highlights geometry as a powerful control parameter for microscale transport, active materials, and soft-matter physics.

Keywords: azopolymer particle, Porcupine-like colloids, microfluidic, Janus particle, Micro/nanoparticle morphology control

**Introduction**

Polymeric particles have become remarkably important across a diverse range of fields, from pharmaceuticals and advanced materials to personal care and medical imaging [1]. Their interest lies in their adaptability, extending foundational research areas, such as microfluidics and nanotechnology. A crucial point to highlight is the significant influence of particle shape on functionality parameters such as microswimmer control, liquid flow dynamics, and motion behavior [2]. These parameters show a highly shape-dependence on the viability of the results. Historically, producing polymer particles with precise and varied shapes posed considerable challenges, resulting in a predominance of simple, spherical forms. However, recent technological advancements have led to new strategies for fabricating polymeric particles with controlled, non-spherical geometries [3].

Among polymer nanoparticles, polymeric Janus nanoparticles are colloidal objects whose surfaces are deliberately divided into two (or more) chemically or physically distinct regions. In the simplest case, each hemisphere is composed of a different polymer or carries orthogonal functional groups, endowing the particle with an intrinsic directionality [4]. This broken symmetry translates into emergent behaviors that are inaccessible to isotropic spheres, positioning Janus nanoparticles at the intersection of advanced soft-matter physics and programmable self-assembly.

Synthesis strategies have evolved rapidly over the past two decades. Seeded emulsion polymerization and phase-segregating block-copolymer micellization were among the earliest routes, but recent researches rely on microfluidic flow focusing [5], electrohydrodynamic co-jetting, and pickering-emulsion masking to tailor particle size (20 nm – 500 nm) and interfacial sharpness [6, 7].

The anisotropic architecture confers distinctive physicochemical characteristics. Under an external field (electric, magnetic, or acoustic) Janus particles experience a net torque rather than purely translational forces, enabling controlled orientation and propulsion like microscopic 'sailboats' [8]. Interfacial tension gradients across the two faces give rise to self-propelling Marangoni flows, while selective functionalization allows orthogonal bioconjugation—antibodies on one side, imaging dyes or drugs on the other—yielding multifunctional nanocarriers that can navigate complex biological materials. Furthermore, the directional dipole in surface energy facilitates hierarchical assembly into one-dimensional chains, two-dimensional patchy crystals, or complex colloidal molecules, expanding the toolbox for bottom-up materials fabrication.

Applications are correspondingly diverse. Photonic crystals derived from self-assembled Janus colloids exhibit angle-independent structural coloration, promising eco-friendly pigments [9]. Despite these advances, challenges remain and new shapes of polymer nanoparticles are still in demand with active stimuli for changing the shape. Azopolymer Janus nanoparticles merge the light-responsive power of azobenzene chromophores with the directional functionality characteristic of Janus architectures. Since the first reports demonstrating their fabrication [10], research has progressed along different axes mainly synthetic methodology, photo-actuated mechanics.

The important point of azopolymers is the reversible trans - cis photo-isomerization. Zhou et al. reported 'Azo-Polymer Janus Particles and Their Photoinduced Symmetry-Breaking

Deformation' where methacrylate-based azopolymer (PAZO-ADMA) was combined with PMMA to form micron-scale Janus particles. Under polarized light, only the azopolymer hemisphere deforms, leading to controllable, reversible symmetry breaking and anisotropic shape changes [11]. C. Liao et al. extended this concept to "Photodeformable Azo Polymer Janus Particles obtained upon nonsolvent-induced phase separation and asynchronous aggregation," where phase separation during particle formation yields azopolymer-rich and azopolymer-poor domains [12].

In all these studies janus nanoparticles made with azopolymer have only two sides. Recently, the possibility to have multiple side on the same sphere was demonstrated [13]. Preparation of raspberry-like silica nanoparticles is fairly easy and well described in the literature [14]. The possibility to fabricate high surface roughness like raspberry-like particles was demonstrated [15]. C. Hsu et al. gave the first demonstration of Janus and strawberry particles using azo molecular glass, photodeformable strawberry-like particles [16]. They presented a simple one-step fabrication method with microparticle's morphology controlled by diffusion and solvent quality effects. Strawberry-like particles (with EG) have partial phase separation leading to small domains (patches). The Azo domains are distributed on PDMS matrix. The study aims to fabricate and control complex microparticles (Janus and strawberry-like) made of:

- Azo molecular glass (IA-Chol) is photoresponsive,
- PDMS oligomer (H2pdca-PDMS) is flexible with low surface energy

and to understand the mechanism of their formation and behavior.

The slow solvent evaporation in droplets creates phase separation between PDMS and azo glass.

Adding Ethylene Glycol blocks gives a full segregation with switches morphology from Janus to strawberry-like. The azo component enables light-controlled elongation and filaments only on the supporting particle. However, the produced filaments are only on the basis of the particle in contact with the substrate and not directly observed on the particle itself with photoinduced patches.

The new multi-janus azopolymer nanoparticles presented in this study can be photo-writable on the different faces and due to the observed hairy surface is named porcupine nanoparticle. Different morphologies could be accessible through a remote process. Finally, we show through Brownian and Jeffery-flow simulations that these photoinduced geometries have profound consequences for hydrodynamic behavior—including anisotropic diffusion, reduced rotational randomization, and enhanced shear-induced dispersion—highlighting geometry as a key regulator of microscale transport.

Materials and methods

The synthesis of porcupine nanoparticles was done using a modified emulsion polymerization technique to ensure the formation of a core-shell structure. The core consists of PMMA, while the shell is made up of azopolymer fragments. The synthesis process is outlined as follows:

The porcupine nanoparticles were fabricated through a self-assembly approach by adding water to a solution of the polymer in chloroform.

Polymers: 20 mg of azobenzene polymer and 20 mg of PMMA ((Aldrich, inherent viscosity ~1.25 dL/g) were dissolved in 2 ml of chloroform. 1 ml of the obtained solution was added carefully into the bottom of aqueous solution of poly(vinyl alcohol (2,5 wt%, 30 ml) under moderate stirring at an appropriate rate (1 ml/min). Then, the mixing speed was increased to ca. 450 rpm to fully disperse the organic phase. In order to slowly volatilizing chloroform, the particles in poly(vinyl alcohol) dispersions were finally formed by slow evaporation of chloroform, which was carried out under the dust-free ambient condition at room temperature with an evaporating rate around 0.1 mL/h for 24 h.

Then, the suspension was centrifuged and washed with water several times.

By adjusting the feed ratio between the azopolymer and PMMA (0.5:0.5) the current morphology of the porcupine nanoparticles were obtained. The resulting particles underwent washing, centrifugation, and drying before further characterization.

The characterization of Composite particles with protruding head composed of azopolymer attached to PMMA body was performed using a scanning electron microscope (SEM - JOEL 6301F), operated at an accelerating voltage of 3 kV.

For photo-switching experiments, the optical setup consists of a horizontally polarized diode pumped solid-state laser beam operating at λ = 473 nm with 100 mW maximum power. The laser excites the azopolymer close to its absorption maximum. Absorbance at the 473 nm working wavelength is 0.64. The sample is set perpendicular to the incident laser beam. The laser beam impinging on the polymer sample is collimated into a plane wave and its size is adjusted with a Kepler-type afocal system. The polarization and direction of the polarization of the beam is controlled via quarter/half wave plate. The power density at the sample location is attenuated down to 0.7 W/cm$^2$.

Results

The morphology of the synthesized polymeric particles was predominantly spherical, as the interfacial tension between the particles and the surrounding medium was minimized. Interestingly, the newly synthesized azospheres resemble a buckyball or soccer ball, characterized by a PMMA core surrounded by azopolymer patches rather than being hollow (Fig. 1). The particle distribution after fabrication is shown in Figure SI1. A droplet of the colloidal solution is deposited on a glass substrate. In diameter, the particle have a range of diameters from less than 10 nm to more than 50 nm. The particle fabrication process allows them to be reproduced with the same size distribution (Fig. SI2) and the same number of patches. The average size of the particles is 40 ± 1 nm. The number of patches is estimated to be 1 patch in 27 nm$^2$. The patch has a semi-spherical shape with a base of 5 ± 1 nm and a height of 1 ± 0.5 nm (fig. SI3).

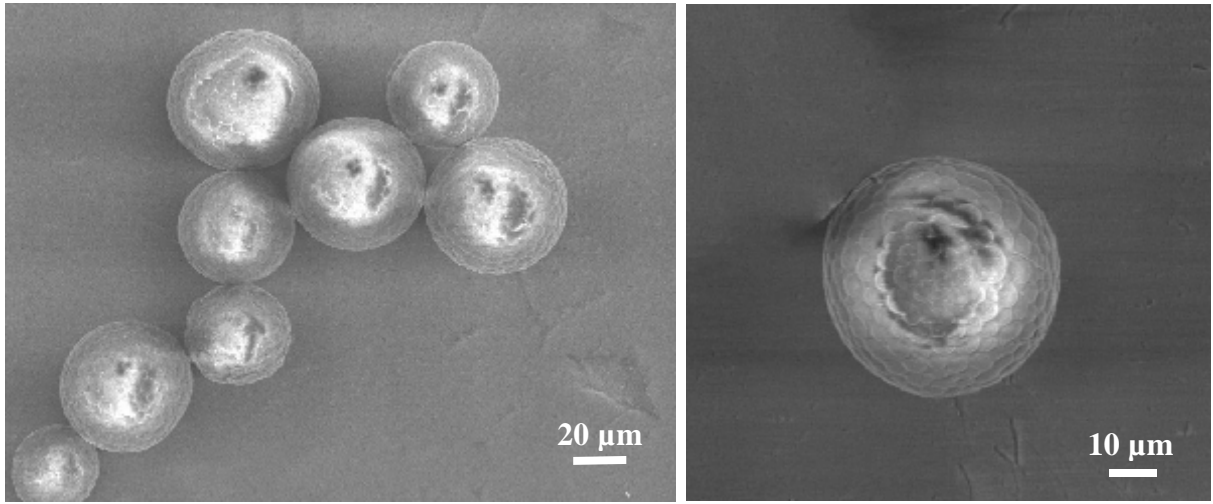

Fig. 1: Fabrication of PMMA spheres with compartmentalized layer as patch of azopolymer on the external surface.

Due to directional photo-fluidization of azopolymers, shell compartimentalized layer of azopolymer can be light manipulated to obtain unique shapes, which can hardly be obtained by other methods [17]. To investigate this possibility, azospheres were subjected to laser irradiation. Upon laser illumination, *trans-cis* photoisomerization occur at each fragment of the azopolymer on the surface of the core PMMA. The photo-fluidization of azopolymer leads to isomerization results in a mass transport of azopolymers in the direction of the polarization of the laser [18]. A TM polarization (vertical polarization) of the laser was first sent to illuminate the new nanoparticles (fig. 2a). The photo-deformation occurs at each fragment of the azopolymer surrounding the core PMMA, resulting in a very unique shape for the initial azospheres. This deformation reaches a saturation point in 15 minutes. After 15 minutes of irradiation time, the final particles look like a porcupine and is shown in the figures 2d and 2e.

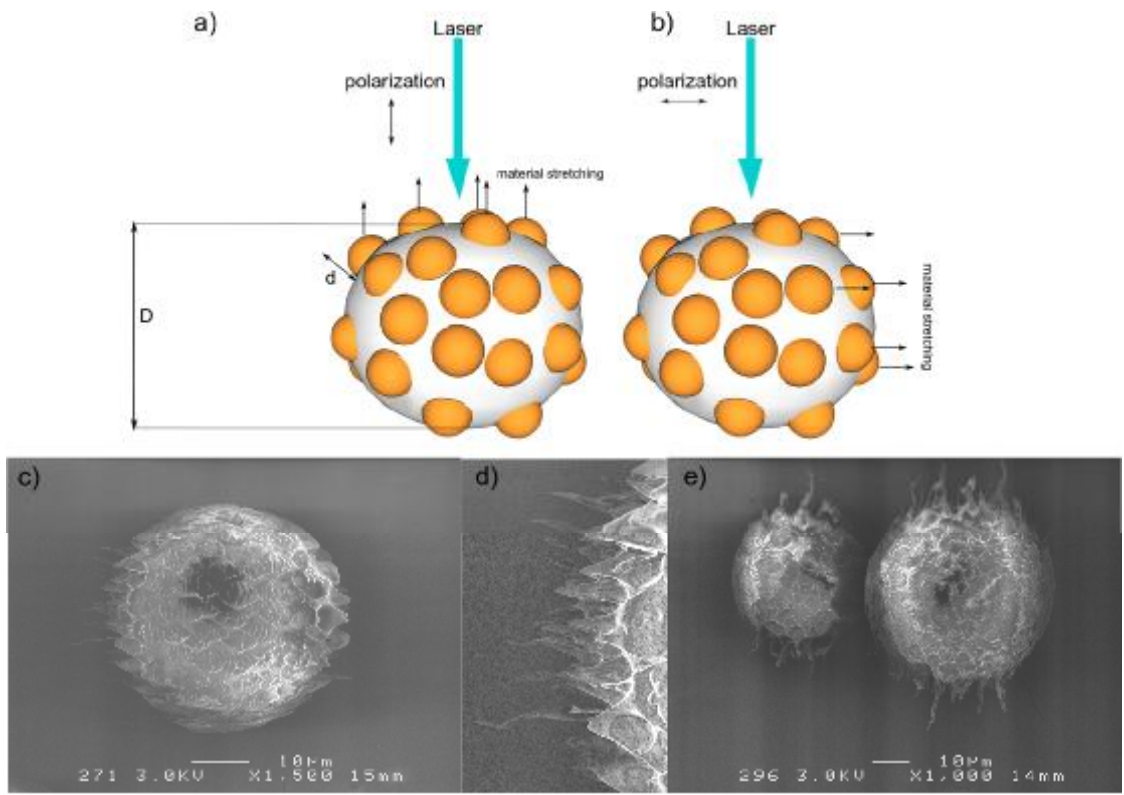

Fig. 2: a) Laser illlumination of new azospheres with a vertical polarization (TM), b) a horizontal polarization (TE). c) The direction of the material stretching by photoisomerization is done in the direction of the laser polarization when the polarization is horizontal or d) details of the filaments observed after illumination by the laser, e) stretching in the vertical direction.

Each patch was balancing between its limitation for being bound to the core and the process for moving along the polarizations direction. Unable to break free of the core PMMA, the azopolymers contained in the fragments started stretching in the direction of polarization and become spikes (fig. 2d). The semi-spherical patches on the main supporting PMMA particle are stretched on a cone shape until a point and then transform in filament. The process is viewed as the following sequence: the light illuminating the patch stretches it in a cone shape. The cone then confines light, focusing more light and giving an amplification of the photofluidization. A filament is then induced at the tip of the cone.

As it has been pointed out in the literature, photofluidization of azopolymer is a polarization dependent phenomenon [19]. To observe the validity of such polarization dependency for these new particles the direction of the polarization of the laser beam was changed to TE (horizontal polarization), using a half wave plate (fig. 2b). The sample of nanoparticles subjected to this polarization exhibit the same behavior, but now the deformation of the compartmentalized layers of azopolymer surrounding the PMMA sphere is observed in the vertical direction (Fig.2c). The transformation into a super-cone occurs across the entire surface of the particle in the direction of the laser polarization. The transformation is only visible at the poles of the particle. However, on the two equatorial parts of the particle, only a surface transformation is observed. The filaments and cones created after light illumination with a linear polarization affect around 30 % of the total surface of the particles. Filaments with length in the range of micrometers have been characterized with SEM observations with a dispersion range from 1 mm to 8 mm.

The resulted 'porcupine' or pollen-like particles [20] are a set of new and exotic shaped particles, opening new doors for studying the phenomena of SRG formation and the basic understanding of the light-matter interaction.

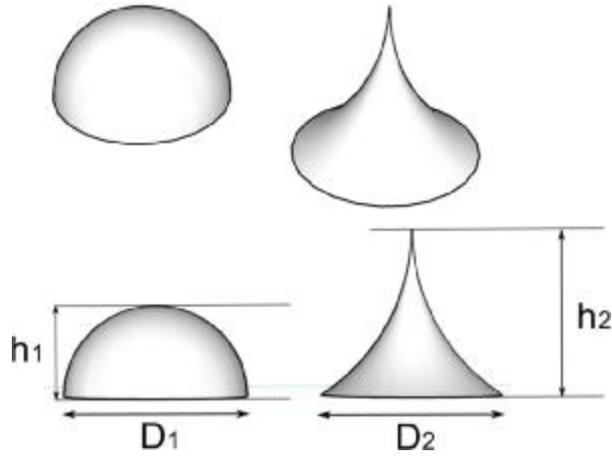

Fig. 3 : Model used in the simulation describing the behavior of the semi-spherical patch on the surface of the particle with $D_1 = D_2 = 5$ mm, $h_1 = 1$ mm, $h_2 = 4$ mm.

The photoinduced effect on the hemispherical azopolymer elements is modelized with a computing software (MATLAB). To this end, it is considered that the volume of the patch is constant and corresponds to the cap shown in Figure 3. The final height is $h_2 = 4$ mm. We consider that the profile of the cone generated as a function of height follows a nonlinear profile of the form:

$$z_{final}(r) = h_2\left(1 - \left(\frac{r}{a_2}\right)^p\right) \text{ with } 0 \leq r \leq a_2$$

with an exponent p of the form $0 < p < 1$. $p = 0.4$ is taken, which allows to fit the shape of the cone obtained experimentally. For $p < 1$, the slope near $r = 0$ becomes very large (almost vertical), leading to a very fine tip close to a filament shape.

The base radius $D_2$ of the 'super-cone' is chosen so as to have exactly the same volume $V_0$ and $h_2 = 4$ mm:

$$V_0 = 2\pi \int_0^{a_2} r z_{final}(r) dr = \pi h_2 a_2^2 \frac{p}{p+2} \Rightarrow a_2 = \sqrt{\frac{V_0(p+2)}{\pi h_2 p}}$$

As a function of time, the hemisphere and the filament cone are superposed, then rescaled to keep $V = V_0$. At the final time, the rescaling factor is 1 and the final height is reached.

The results are presented in the figure 4.

The time is normalized to the final time. The figure 4a present the evolution of the azopolymer patch as a function of time starting from the semispherical shape to a filament or 'super-cone'. The volume is kept constant and the photoinduced phenomenon creates mass transport in a single direction, that of the laser polarization. To produce the filament or super-cone, the surface hollows out and the volume stretches upwards. Figure 4b shows the hollowing out of the surface as a function of time. The radius of curvature of the surface is modeled as a nonlinear function. The angle of curvature is plotted in Figure 4c as a function of time. The evolution is almost linear. It is assumed that the change in shape of the cone is very rapid at the beginning of illumination and slows down as the shape approaches that of the filament or super-cone. In this case, the photoinduced phenomenon slows down and saturates because there is less material to transport.

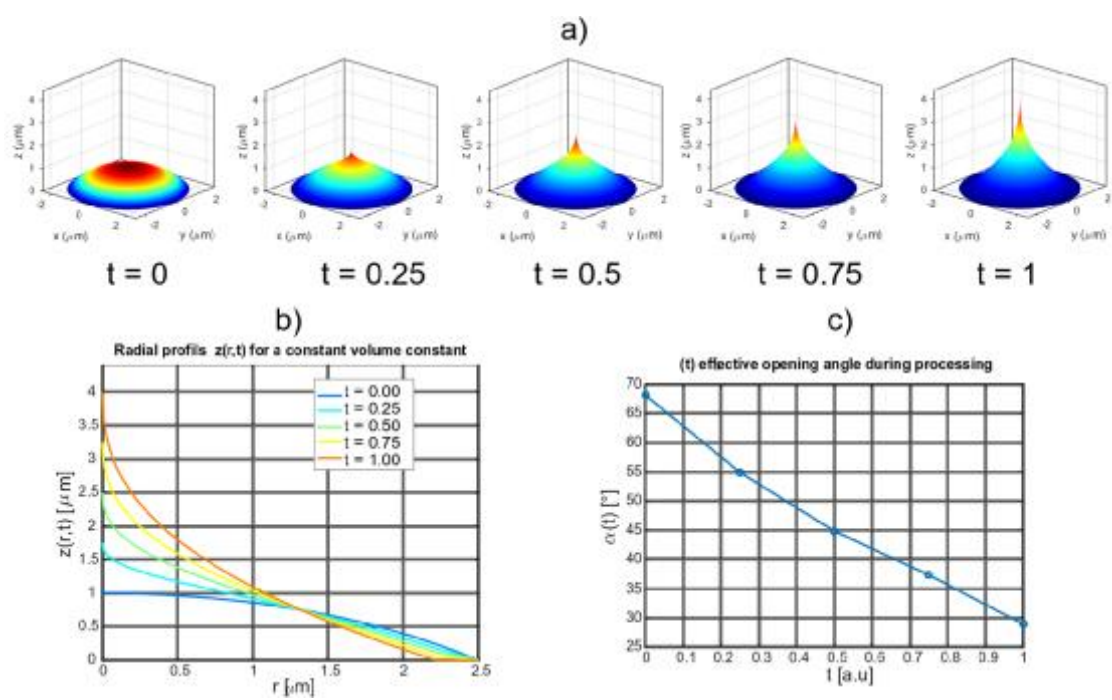

Fig. 4: a) Simulation of evolution of the semi-spherical patch on the azopheres at different times. The time is normalized to the final saturation of the photoinduced phenomenon, b) variation of the patch's surface as a function of time, c) the angle of the cone is changed during illumination.

Following the further possibilities of surface deformations and new shapes using light, the polarization of the laser was changed to circular polarization, with a quarter wave plate. Initially patchy spherical particles containing the azopolymer patches on the surface were illuminated by circular polarization for 15 minutes. The final structure of the particles is shown in Figure 5. These particles resemble a sea pineapple. Under illumination with a circular polarization, there is no any preferred direction of the movement for azopolymers. However, through the *trans-cis* isomerization and movements of the azopolymers it was observed bump shapes clearly visible on the particles resembling a seed or sea pineapple.

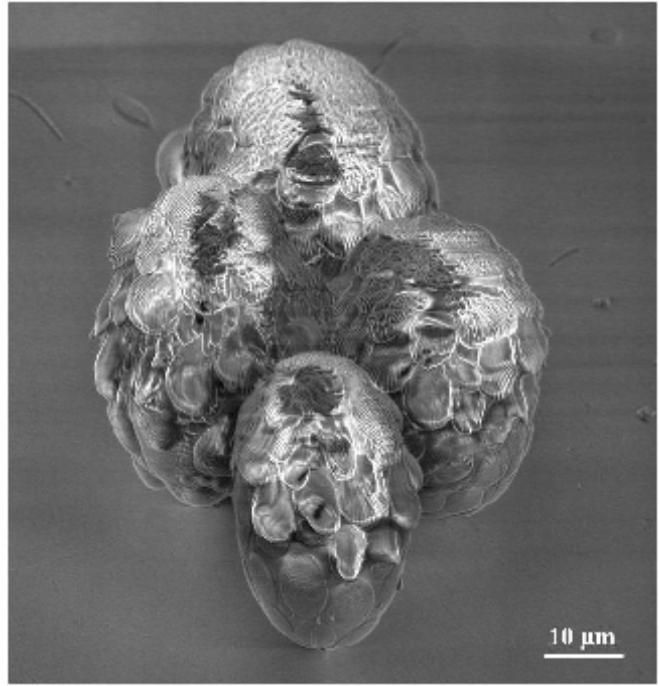

Fig. 5: Illumination of the new azopolymer microspheres with a circular polarization.

Such similar particles were recently synthesized through a complicated three step process. First, highly cross-linked seed particle dispersion was synthesized by emulsifier-free emulsion polymerization with acrylic acid as co-monomer for the formation of surface carboxylic groups. Then, a successive growth scheme was applied to the seeds by swelling the particles with monomer droplets, followed by polymerization. The sea pineapple-shaped particles could be produced by adjusting the amount of monomer during the swelling step of the third growth [21]. Unlike the mentioned process, particles presented in the study using laser illumination were obtained faster and with less effort after fabrication with the result of the mass transport induced by the light as stimulus.

As it was demonstrated in previous researches, inscription of gratings in the surface if azopolymer thin films are done not only by an interference pattern of a double beam exposure [22], but also through a single beam experiment [23]. Following the method of SRG inscription, a sample of these azopolymer particles were subjected to an interference pattern. The laser beam was spatially filtered, collimated and then split by a beam splitter (BS) giving two beams of equal intensity. The two beams, after reflecting from a mirror, are recombined to form an interference pattern and were incident on the sample stage. As can be seen from Fig.6a-b, upon exposure to the interference pattern a surface grating is inscribed on fragments of azopolymer on the particles during the first 5 minutes of irradiation (Fig. 6a).

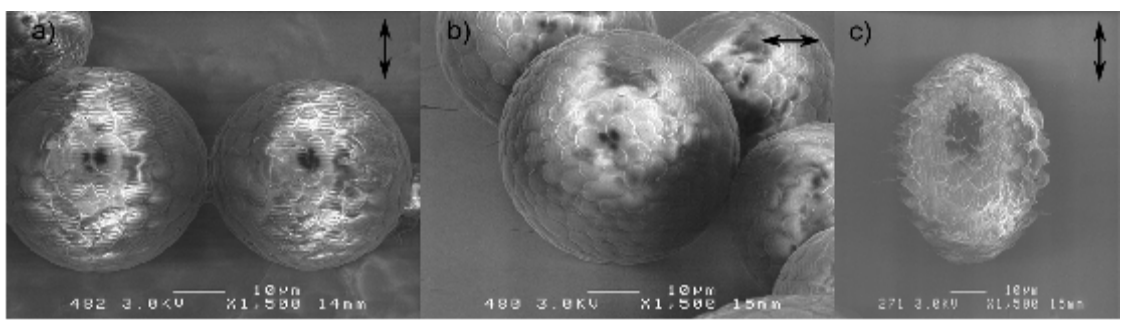

Fig. 6: laser illumination of porcupine nanoparticles with interference patterns. a) the fringes are illuminating with a vertical polarization, b) with a horizontal polarization.

The pitch value (L) for the gratings on these azo fragments (fig. 6a and b) is about 1.3 ± 0.1μm, which is in agreement with the degree of the interference pattern (θ = 10°). By continuous illumination of the samples up to 10 minutes, the grating on the fragments become more dominant and the movement and stretching of the azopolymer start. After 15 minutes of irradiation, the particles are similar to the particles exposed to the single linearly polarized beam, spiked like a porcupine (Fig. 6c).

One of the big advantages of azo incorporated polymers has been their reversibility. Researchers working with these azobenzene based compounds in SRG formation and photo deformation have been using methods such as circular polarization and unpolarized light to reverse the effects applied to the surface and re-do the effects again. Given the fact that the new particles consist in a layer of azopolymer patches, there should be a degree of reversibility for the induced shapes on the particles.

In order to test this reversibility, one of the samples already subjected to linear polarization (Fig. 7a) were irradiated by a beam of circular polarization for 15 minutes. Considering the random directional movement of the azopolymers under such illumination, the spikes on the porcupine particles retreat and change the porcupine shape particles into sea pineapple particles (Fig. 7(b)). The super-cone undergoes photoisomerization, and the previously formed filament flattens under the effect of light-induced mass transport.

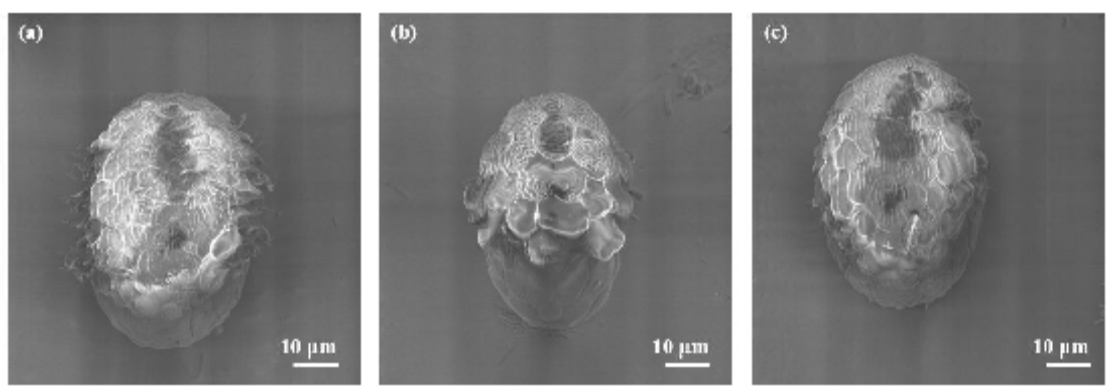

Fig. 7: a) initial particles with super-cone generated by the photoinduced process after illumination with TE polarization, b) the same particle is illuminated with a circular polarization,

Considering the nature of the azopolymers and its photoinduced molecular mass transport, new geometric changes of the particles through a directional polarization could be possible. Hence, the same sample of patchy particles, now with pineapple-shape particles (Fig. 7b), was again irradiated with a horizontally linear polarized laser beam. Once more, the formation of 'porcupine' particles as a result was observed (Fig. 7c).

Beside the unique and exotic shape of these particles under laser illumination, stretching of the azo fragments effect not only the shape of their particles, but also their coverage area. Evaluations from the images show that, for a typical size buckyball particles (with a diameter of 35μm), azo fragments exposed to laser beam stretch about 13 ± 1μm. This value is about 0.4 of the particles diameter, which leads to a particle that, now can cover an elliptical area 1.5 times more than the circular area it was covering before. The figure 8 shows the evolution of the particle after irradiation (fig. 8d). Initially spherical (fig. 8a), the particle is elongated in the direction of the laser polarization and stretch. The stretching is not comparable to a pure azopolymer particle. However, the ellipticity is increased by a factor 0.2 (fig. 8b). A model calculating the elongation of the particle is done and shows the results of the ellipticity after irradiation (fig. 8c, SI4).

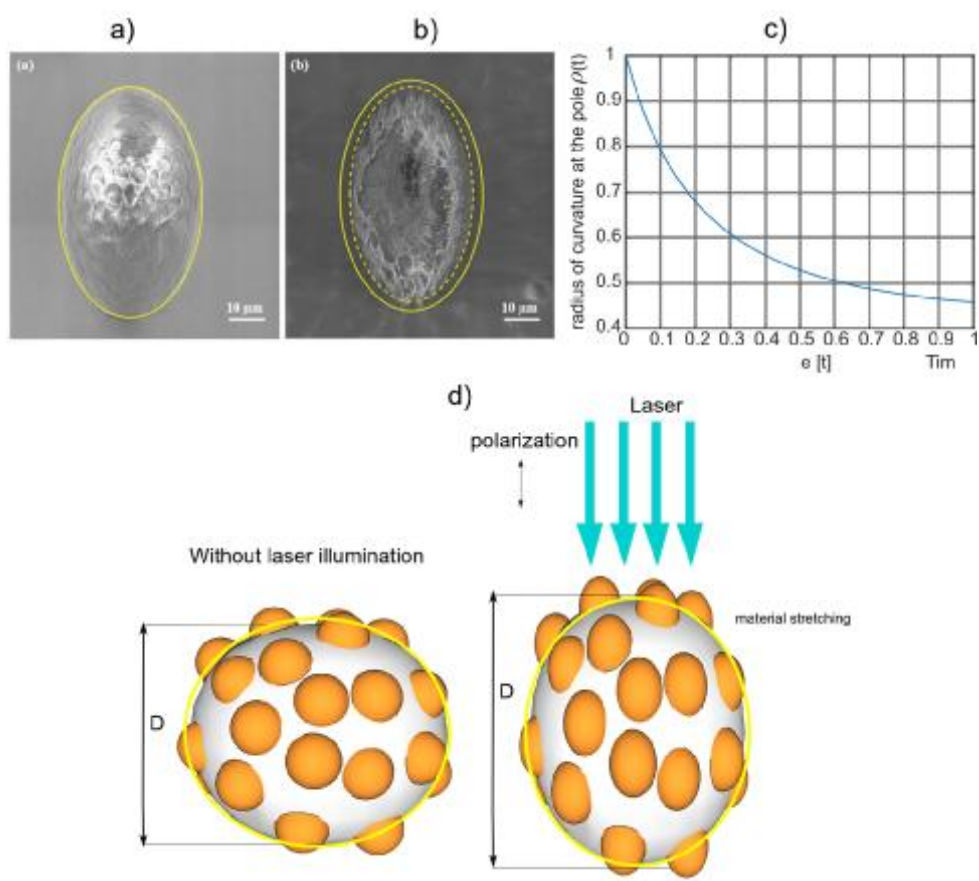

Fig. 8: a) initial particle before irradiation, b) final shape of the particle after illumination, c) evolution of the curvature radius. The photoinduced phenomenon not only shapes the patches in super-cone but stretches the particles itself. The time is normalized to the final.

Availability of polymeric particles with distinct shapes will prove to be a useful tool in uncovering the role of shape in applications such as biological systems. Particles can be used as models for similar- looking objects, like bacteria or pollen grain, not only in biological applications but also in environmental fields to study bacterial migration in soil or water. In another field, non-spherical particles also may open new applications in advanced materials due to their unique scattering and packing properties. These particles also can be used as research tools or probes in rheology or aerodynamics or as sensors in micro-rheology studies. The uses and knowledge gained from particles with defined shapes seem limitless, and the availability of a simple method to make such particles will catalyze many new applications in photonics or soft matter.

In order to compare the impact of these super-cones generated by the photoinduced process after illumination by a linearly polarized laser beam, the behavior of the created particle ("porcupine") is simulated in a liquid medium and compared with a purely spherical particle. Two media are considered: a stable medium and a medium containing a shear flow.

The following geometric is considered:
- a micron-scale particle composed of a spherical core of radius $R_p$ coated with N conical 'scales' of height $h_s$ and base diameter $D_{base}$.
- a spherical particle for comparison.

The half-apex angle of each cone is: $\alpha_s = atan(r_{base}/h_s)$. The scales are distributed symmetrically around the north and south poles within a polar cap of half-angle $\theta_{cap}$ = 30°. Their locations are specified in spherical coordinates ($\theta_k$ = acos($\mu_k$), $\phi_k$) with azimuth uniformly random, and with polar cosine drawn from: $\mu_k = \pm [\mu_{min} + (1 - \mu_{min})\mu_k]$, $\mu_{min} = cos\theta_{cap}$ where $\mu_k \in [0, 1]$ is uniform and the sign ± corresponds to the north/south pole. These geometrical parameters are used only to define the effective hydrodynamic dimensions. No chemical activity is considered here.

The particle is immersed in water at temperature T = 298 K and viscosity η = 10⁻³ Pa·s.

For comparison, a smooth sphere of radius $R_p$ has classical Stokes–Einstein diffusivities [24, 25]:

$$D_{t,sphere} = \frac{k_B T}{6\pi\eta R}$$

$$D_{r,sphere} = \frac{k_B T}{8\pi\eta R_{rot}^3}$$

with $D_t$ translational diffusion and $D_r$ rotational diffusion

To represent the elongated 'pineapple' morphology, effective hydrodynamic radii are introduced [26]:

- Parallel to the symmetry axis: $R_\parallel = R_p$,

- perpendicular to this axis: $R_\perp = R_p + \frac{1}{2}h_s$,

- rotational effective radius: $R_{rot} = R_p + \frac{1}{2}h_s$.

This yields anisotropic translational diffusion:

$$D_\parallel = \frac{k_B T}{6\pi\eta R_\parallel^{eff}}, \quad D_\perp = \frac{k_B T}{6\pi\eta R_\perp^{eff}},$$

and a reduced rotational diffusivity : $D_r = \frac{k_B T}{6\pi\eta \left(R_{rot}^{eff}\right)^2}.$

In the 'smooth sphere' mode of the simulation: $D_\parallel = D_\perp = D_{t,sphere}$ , $D_r = D_{r,sphere}$.
The particle configuration is described i) by the position vector X(t), ii) a unit orientation vector p(t) with $\|p(t)\| = 1$ which defines the diffusion anisotropy.
At each time step, an orthonormal frame (p, e₁, e₂) is constructed with $e_1 \perp p$, $e_2 = p \times e_1$ (fig. SI4).
The orientation of the particles evolves via isotropic rotational diffusion Dp = (DW$_{rot}$ × p), where DW$_{rot}$ is a 3D Wiener increment with covariance $\langle D\Omega_{rot,i}\, D\Omega_{rot,j}\rangle = 2D_r \delta_{ij} dt$. Using the discrete Euler-Maruyama form it gives:

$$p_{n+1} = \frac{p_n + (\Delta\Omega_{rot} \times p_n)}{\|p_n + (\Delta\Omega_{rot} \times p_n)\|} \text{ with } \Delta\Omega_{rot} \sim N(0, 2D_r, dt)$$

The random rotational vector ΔW$_{rot}$ = (ΔW$_x$, ΔW$_y$, ΔW$_z$) is distributed according to a 3-dimensional multivariate normal (Gaussian) distribution, with mean zero and a covariance matrix equal to $2D_r dt$ (**fig SI5**). ΔW$_{rot}$ is the source of the angle diffusion and is due to molecular collision in water. In the diffusion process $\langle(\Delta\theta)^2\rangle = 2D_r dt$
This update enforces unit length at each step.
The particle's position obeys the stochastic differential equation and its position is given by:

$$X(t + \Delta t) = X(t) + U_{flow}\Delta t + \Delta X_\parallel + \Delta X_\perp$$

with $\Delta X_{brown} = \Delta X_\parallel + \Delta X_\perp$

$$dX = U_{flow}dt + \sqrt{2D_\parallel} p\, d\Omega_\parallel + \sqrt{2D_\perp}(e_1 d\Omega_1 + e_2 d\Omega_2)$$

where $d\Omega_\parallel, d\Omega_1, d\Omega_2$ are independent Wiener increments. The position is updated:

$$X_{n+1} = X_n + U_{flow}dt + \sqrt{2D_\parallel dt}\, z_\parallel p + \sqrt{2D_\perp dt}(z_1 e_1 + z_2 e_2)$$

with $z_\parallel$, $z_1$, $z_2$ **are** independent Gaussian random variables, each following a reduced centered normal distribution ($z_\parallel \sim N(0,1), z_1, z_2 \sim N(0,1)$). Finally, The simulation computes the time-dependent mean squared displacement (MSD) with MSD(t) = $\|X(t) - X(0)\|^2$.
The calculation leads to a full 3D anisotropic translational Brownian dynamics scheme. The presented model is purely passive with no chemical activity or surface flux (i.e. no autophoretic or diffusiophoretic mechanism), no hydrodynamic shear or Jeffery rotation, no external torques or forces beyond background flow and Brownian fluctuations. The model isolates the impact of morphology (scales/filaments) on translational and rotational diffusion.
The simulations are performed in an unbounded simple shear flow with identical flow and initial conditions for all particle types. While the external hydrodynamic boundary conditions are unchanged, the effective surface boundary conditions differ between models. The spherical particle satisfies an isotropic no-slip condition and exhibits purely rotational motion dictated by the local vorticity. The ellipsoidal particle follows classical Jeffery dynamics due to its anisotropic shape under no-slip conditions. In contrast, the 'pineapple' particle is not geometrically resolved but modelled through an effective anisotropic surface response, introducing a perturbation to the Jeffery rotation. As a result, differences in dynamics arise solely from the modified surface boundary condition and particle geometry. The particle orientation is initialized with a prescribed unit vector and evolved in time. When Brownian effects are included, rotational diffusion is modelled as an additive stochastic contribution with Gaussian white noise, ensuring isotropic fluctuations superimposed on the deterministic Jeffery motion.

The results are presented in the figure 9.

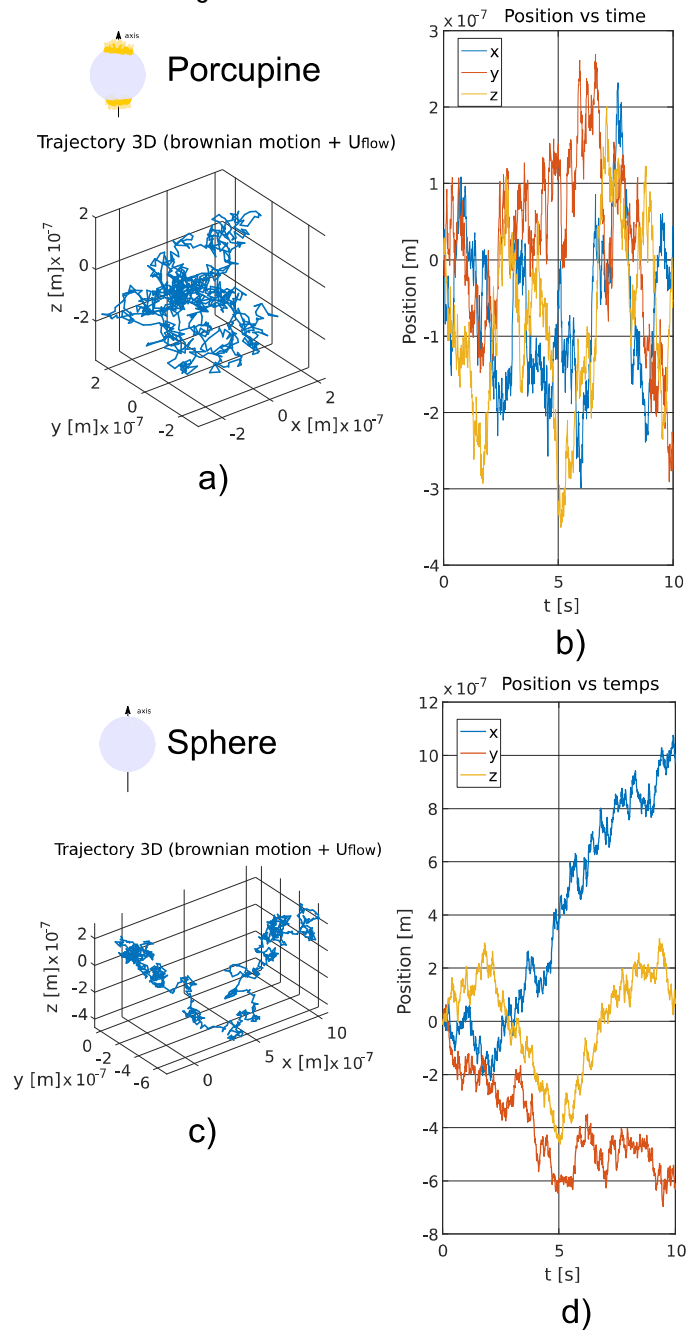

Fig. 9: Comparison of diffusion of the particle with self-induced morphological shape change ('porcupine-like' particle) in water and a pure spherical shape particle with the same size. The particle diameter is 30 nm.

The anisotropic morphology of the 'porcupine-like' particle—considering a spherical core coated with elongated conical scales—induces a marked imbalance between diffusion parallel and perpendicular to the particle's symmetry axis. Because translation along the axis encounters a smaller effective hydrodynamic radius ($R_\parallel = R_p$), the corresponding diffusivity is enhanced ($D_\parallel > D_\perp$), whereas motions in directions normal to this axis is hindered by an enlarged cross-section ($R_\perp = R_p + h_s/2$). This anisotropy has several dynamical consequences. First, the instantaneous Brownian steps become elongated, generating trajectories that show short-time directionality even in the absence of external forces or flows. Second, because rotational diffusion is simultaneously reduced (owing to the larger rotational radius $R_{rot}$), the orientation vector p(t) decorrelates more slowly, allowing the particle to maintain its direction of fastest diffusion for extended periods. The joint effect is a transient regime of superficially persistent or 'quasi-ballistic' Brownian motion, visible in the MSD as an early-time slope higher than that of an equivalent smooth sphere (fig. SI7). At longer times, full rotational randomization restores isotropy in the ensemble average, but with a larger effective diffusivity than the spherical particle because the particle spends more time moving along directions where diffusion is intrinsically faster. This behavior is characteristic of anisotropic colloids such as rods, ellipsoids, and biologically inspired particles [26], and here it emerges purely from the hierarchical spiky geometry of the porcupine particle. Importantly, the model isolates morphology as the sole source of anisotropy: no chemical activity, flow alignment, or hydrodynamic torques are present. Thus, the observed enhancement and transient directionality in the MSD provide a clear quantitative signature of how geometry alone—and specifically the condition $D_\parallel > D_\perp$ - modulates microscale stochastic transport in water.

The figure 10 shows a direct comparison between the Brownian motion of a porcupine-like particle (spherical core + outward filaments, anisotropic diffusion), and a smooth sphere of the same core size (isotropic diffusion). Several characteristic behaviors emerge. A stronger directional drift in the porcupine particle is observed (fig. 10a).

Moreover, with the 3D trajectory it is observed more persistent, elongated path for the porcupine-like particle. The 3D plot shows that the sphere particle executes a compact, isotropic random walk, with no long-range directional bias. The porcupine-like particle's trajectory is much more extended and visually elongated. This occurs because the porcupine has $\langle (\Delta r)^2 \rangle_{porcupine} = \langle (\Delta r)^2 \rangle_{sphere}$ over the same time due to its larger effective diffusion along its axis. Slow orientation randomization enhances anisotropy. The porcupine's long filaments reduce the effective rotational radius $D_r \sim 1/R_{rot}^3$.

The orientation *p(t)* of the porcupine particle changes slowly. The particle keeps 'pointing' in nearly the same direction. This effect produces an extended, semi-ballistic regime of motion. The smooth sphere, by contrast, reorients almost instantly and therefore averages out any directional bias.

In the time-series plot (fig. 10b), one component of the sphere trajectory grows significantly in magnitude over ≈ 10 s, while the porcupine-like particle's components fluctuate around zero with no preferred direction. This is a hallmark of anisotropic diffusion where $D_\parallel > D_\perp$ and the particle's orientation p(t) persists for long periods because rotational diffusion is reduced by the filaments. Brownian steps are larger along the axis of lowest drag (the porcupine's axis). Because *p(t)* rotates more slowly (low $D_r$), the porcupine-like particle continues moving preferentially in the same direction for seconds. This generates an apparent directional drift even though no force is applied. This 'drift' is not a force but a persistent random walk, typical of anisotropic colloids (rods, ellipsoids, bacteria-sized filaments).

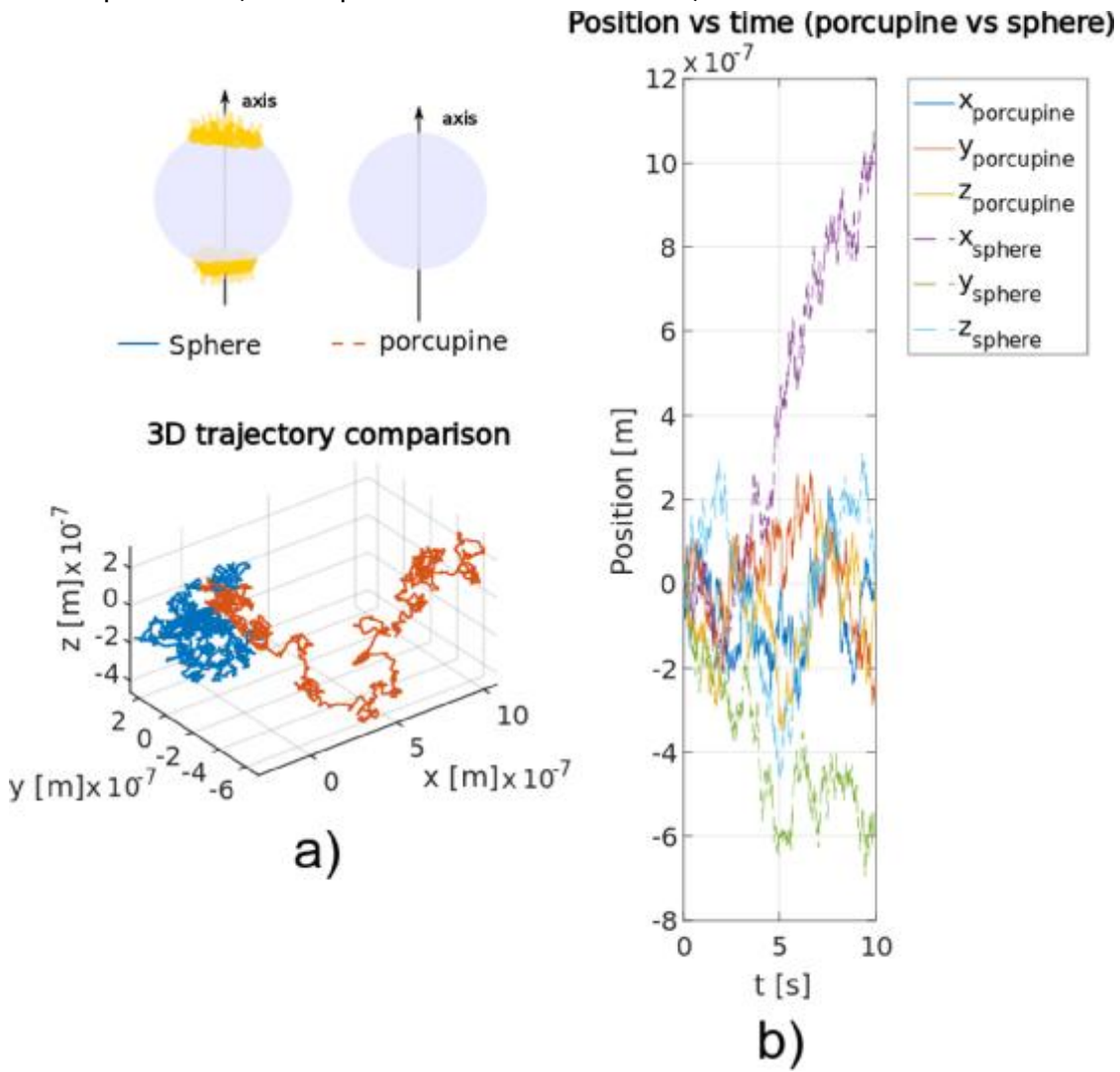

Fig. 10: Comparison of the geometry influence of the particle on its trajectory in the liquid.

The porcupine particle undergoes enhanced, directionally persistent diffusion, moving farther and more coherently because it has a lower hydrodynamic drag along its axis and a slower rotational diffusion. The sphere exhibits classical isotropic Brownian motion, with no long-time directional behavior. The comparison illustrates how geometry alone—without any chemical activity—can produce significant anisotropy in microscale transport.

In a second model orientation dynamics with Jeffery, Brownian and Gravity effects is considered. p(t) is the unit orientation vector of the particle. Its evolution combines Jeffery hydrodynamic torque [28] with a simple shear:

$$u = (\dot{\gamma}y, 0, 0)$$

$$G = \nabla u = \begin{pmatrix} 0 & \dot{\gamma} & 0 \\ 0 & 0 & 0 \\ 0 & 0 & 0 \end{pmatrix}$$

with $E = \frac{1}{2}(G + G^T)$ the rate of strain of the flow, $\Omega = \frac{1}{2}(G - G^T)$, the velocity field of the fluid in a simple shear (rate of rotation of the flow) with x = flow direction, y = velocity gradient direction, z = neutral direction. G = ∇u (the velocity gradient tensor).

G measures how velocity varies in space. E and Ω are constructed from G. E is the strain rate tensor (symmetric part). This is the symmetric part of G. It represents the pure strain of the fluid (stretching/compression). This term enters into the Jeffery equation. Ω is the vorticity tensor (antisymmetric part). This is the antisymmetric part of G. It corresponds to the local solid rotation of the fluid (vorticity). In Jeffery's equation, the term Ω.p tends to rotate the particle like a small fluid element [29, 30].

The Jeffery equation for a particle with effective aspect ratio r is [31]:

$$\dot{p}_{j\,ff} = \Omega p + \lambda_J(E - (p^T E)p) \text{ avec } \lambda_J = \frac{r^2-}{r^2+} \text{ and } r = 1 + \frac{h_s}{R_p}.$$

where $\lambda_J$ is the Jeffrey parameter [31].

If the center of mass is slightly displaced, gravity introduces a restoring torque proportional to:

$$\dot{p}_{g\,a} = \alpha_{g\,a}\,(p \times \hat{g})$$

where $\hat{g}$ is the unit gravity vector and $\alpha_{grav}$ sets the torque intensity. The rotational diffusion is implemented through: $dW_{rot}$ a random variable as previously defined and with stochastic increment: $dp_{sto} = dW_{rot} \times p$.

Combining the three contributions $\dot{p} = \dot{p}_{j\,ff} + \dot{p}_{g\,a} + stochastic\ term$ with the discrete Euler-Maruyama upgrading for each position, we have:

$$p_{n\,1} = \frac{p_n + dt\dot{p}_{j\,ff} + p_n + dt\dot{p}_{sto}}{\|p_n + dt\dot{p}_{j\,ff} + p_n + dt\dot{p}_{sto}\|}$$

Considering a translational motion with anisotropic brownian dynamics and shear advection, at each step, an orthonormal frame (p, $e_1$, $e_2$) is built as previously with $e_1 \perp p$ and $e_2 = p \times e_1$
The anisotropic Brownian displacement is:

$$dX_{aniso} = \sqrt{2D_\parallel dt}z_\parallel p + \sqrt{2D_\perp dt}(z_1 e_1 + z_2 e_2)$$

with the same random variable previously described $z_1$, $z_2$, $z_3$ and p(t) the orientation, anisotropic diffusivities $D_\parallel$, $D_\perp$.

The shear flow contributes deterministic advection: $U_{shear} = (\dot{\gamma}, 0, 0)$. The full position (trajectory) updated is:

$$X_{n\,1} = X_n + dtU_{shear} + dX_{aniso}$$

The results are plotted in the figure 11 for the two particles.

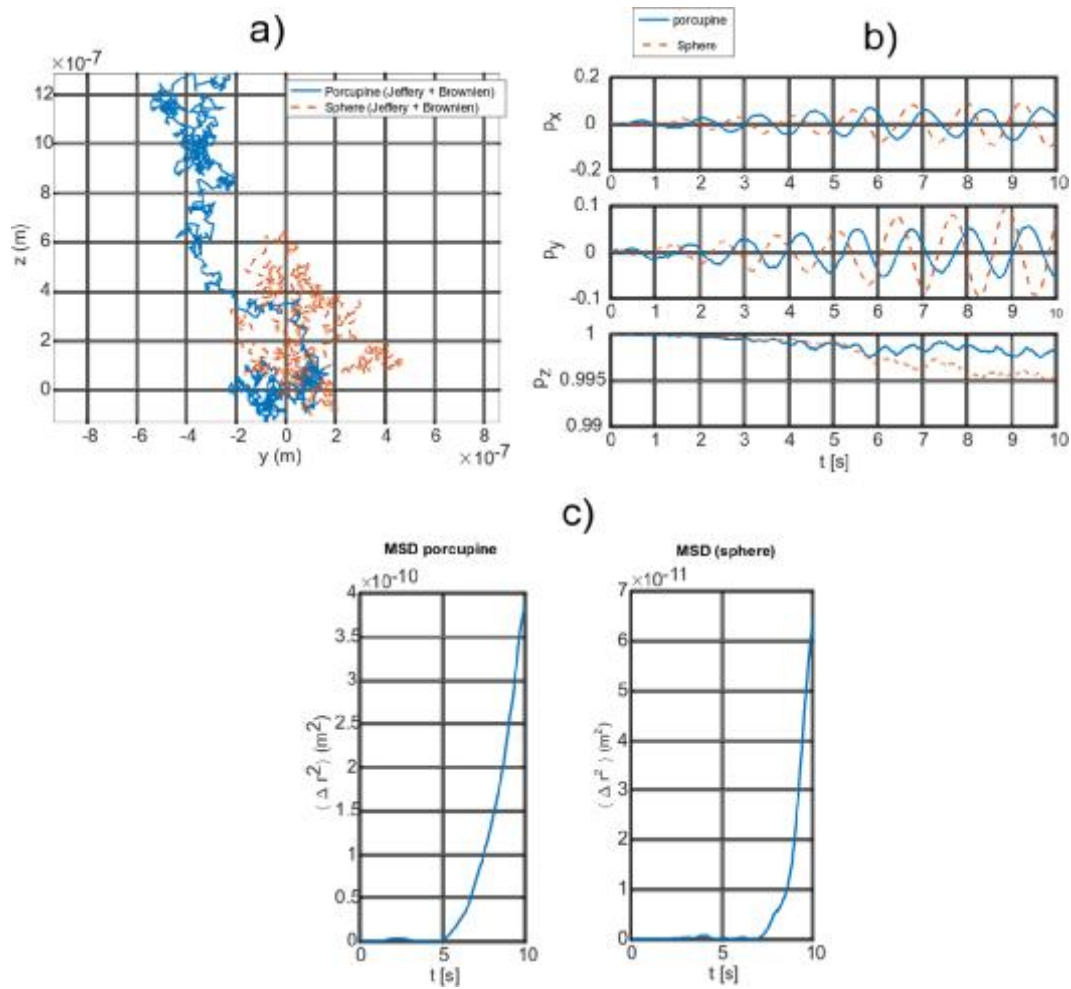

Fig. 11: Comparison of the behavior of the two particles in a fluid with shear flow. a) trajectories of the two particles with b) the rotation components as a function of time, and c) the mean square displacement (MSD).

The trajectories in the (y, z) plane (fig. 11a) shows that for the 'porcupine' particle, the trajectory is much more extended and 'organized' in large loops along z. This is because the anisotropic particle (aspect ratio r > 1) follows Jeffery orbits in the shear. Its axis p(t) rotates periodically in the shear plane, and since diffusion is faster along p ($D_\parallel > D_\perp$), each Jeffery orbit results in a greater net displacement in certain directions.

For the sphere, the trajectory remains compact. For a sphere r = 1 ⇒ $\lambda_J$ = 0 and the Jeffery term is reduced to the simple solid rotation of the fluid, which has no effect on isotropic diffusion. It is observed mainly Brownian motion + shear advection, without geometric amplification.

The orientation p(t) = ($p_x$, $p_y$, $p_z$) is presented in the figure 11b. For the 'porcupine' particle the $p_x$ and $p_y$ components exhibit quasi-periodic oscillations, typical of Jeffrey orbits of an elongated particle in simple shear. $p_z$ remains close to 1 but shows small modulations, a sign that the particle is tilting regularly.

For the sphere, the components remain almost constant (or slightly noisy). Since $\lambda_J$ = 0, Jeffery's

E-dependent term disappears, so the sphere does not 'feel' the orientation in the shear: no orbits is observed, only rotational noise.

The figure 11c compare the MSD for 'porcupine' vs sphere particle. The MSD of the porcupine is almost an order of magnitude greater than that of the sphere at the same time. Jeffery and anisotropy **lead to a** strongly amplified dispersion.

The MSD of the sphere mainly reflects classical Brownian diffusion in a shear flow, with no shape-related gain.

The combination of elongated shape, Jeffery and $D_\parallel > D_\perp$ creates orientation orbits that transform shear into a super-dispersion mechanism for the porcupine particle.

The sphere, for which $\lambda_J = 0$ and $D_\parallel = D_\perp$, does not benefit from any of these geometric resonances and remains much less mobile.

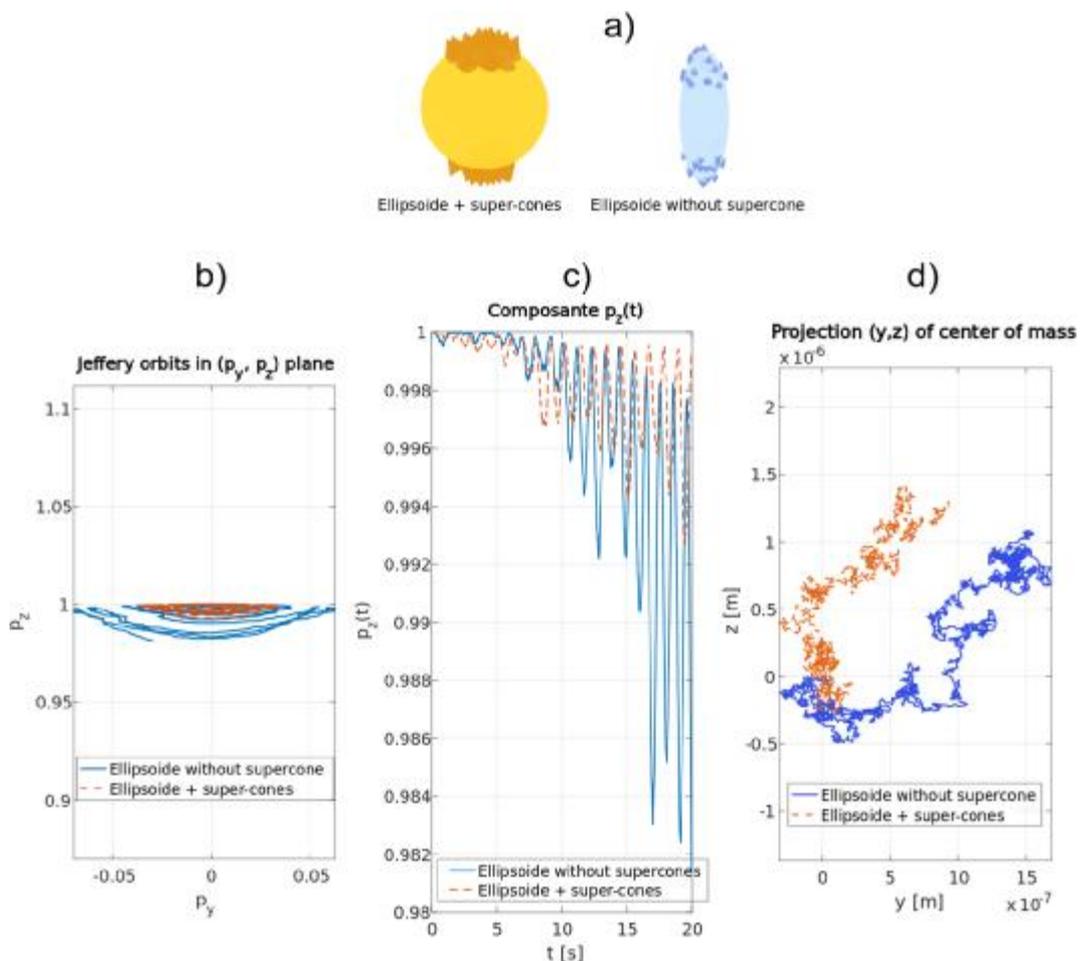

Fig. 12: Comparison of the behavior of the two ellipsoid particles with and without photoinduced super-cones in a fluid with shear flow. a) Jeffery orbits of the two particles b) the rotation component p(z) as a function of time, and c) projection of the center of mass showing the behavior of the two particles as a function of time.

In order to test the influence of Super-Cones on Hydrodynamic Response, calculations were done for an ellipsoid shape particle with super-cones generated by the photoinduced response of the patches in the poles (fig. 12). Only the azopolymer patches can be observed in the direction of the polarization. The other patches are elongated along the surface. The results are compared with the same particle with generated super-cone. In the present case the shape of the particle is due to the linear polarized laser illuminated the particle and a shape deformation of the whole particle induced by the total patches. Figure 12 compares the orientational dynamics of an ellipsoidal particle with and without super-cones under identical shear flow conditions. The Jeffery orbits projected in the ($p_y$, $p_z$) plane show that the smooth ellipsoid undergoes wider periodic excursions, whereas the super-cone particle exhibits noticeably compressed orbits with reduced angular amplitude. This reduction is confirmed in the time evolution of $p_z(t)$, where the smooth ellipsoid displays large oscillations while the super-cone particle remains more closely aligned with the flow, with smaller and more regular modulations. With Brownian noise, the translation of the center of mass in the (y, z) plane strongly differs between the two cases: the smooth ellipsoid shows broader lateral wandering, while the super-cone particle follows a significantly more confined trajectory. Because rotational fluctuations are smaller, the coupling between rotation and translation is weakened, leading to a measurable reduction in Brownian wandering. Overall, the simulations show that adding super-cones alters both the deterministic Jeffery rotation and the stochastic translational diffusion, leading to more stable orientation and reduced spatial dispersion. The presence of super-cones at the poles modifies the effective hydrodynamic resistance of the particle and therefore changes its behavior in shear flow. The anisotropic roughness increases rotational drag near the poles, which suppresses large Jeffery excursions and keeps the particle preferentially aligned with the flow direction. These results indicate that even modest geometric modifications—restricted to the poles—can substantially shift the orientation statistics and transport properties of anisotropic particles in viscous shear flows.

Conclusion

In this work, a straightforward optical method to fabricate complex azopolymer–PMMA composite nanoparticles whose shapes can be reversibly sculpted with light is presented. The patchy azopolymer surface enables localized, polarization-controlled photo-fluidization, allowing spherical particles to transform into filaments, spikes, or isotropic bumps. This bottom-up strategy offers access to shapes called porcupine and sea-pineapple morphologies—that are difficult or impossible to obtain using conventional multi-step synthetic routes. A nonlinear geometric model accurately reproduces the experimentally observed patch-to-cone transformation and captures the rapid–then-saturating dynamics of photoinduced mass transport.

Beyond shape control, we show that these optical deformations drastically modify the particles' hydrodynamic behavior. The porcupine geometry generates strong anisotropic Brownian diffusion, slow rotational decorrelation, and Jeffery-orbit-enhanced dispersion in shear flow. Even modest geometric modifications, such as photoinduced super-cones at the poles of an ellipsoid, significantly alter orientation statistics and translational transport. Together, these findings reveal geometry as a powerful and tunable element of colloidal design. The ease of fabrication, reversibility, and dynamical consequences of these new particles open promising perspectives for responsive materials, optical micromanipulation, biological model systems, and next-generation micro- and nanoscale transport studies. One possible method to ensure an efficient homogeneous illumination is to do the illumination in a microfluidic channel with a periodic sorting of micro-particles and a control of the flow.

The recent review on nanoparticle-mediated regulation of cell death [33] emphasizes how engineered nanostructures, through precise control of morphology, surface chemistry, and stimuli responsiveness, can modulate apoptosis, ferroptosis, and other biological pathways for therapeutic purposes. In parallel, the phage-lysin functionalized red bacterial microparticles demonstrate how biologically programmed surface recognition combined with visible optical readout enables rapid point-of-care diagnostics [34]. Our light-sculpted azopolymer porcupine particles conceptually bridge these two directions: they provide a remotely reconfigurable micro-platform whose geometry, surface anisotropy, and hydrodynamic response can be dynamically programmed by light polarization. Beyond their fundamental soft-matter interest, these particles could evolve into multifunctional bioactive carriers by (i) grafting targeting ligands or lysin-derived recognition domains onto the azopolymer spikes to create geometry-amplified biosensors, (ii) using their polarization-controlled anisotropic transport for enhanced microfluidic sorting or shear-triggered drug release, and (iii) exploiting their high surface curvature and patch multiplicity as platforms for multivalent biochemical interactions. In this perspective, the porcupine morphology is not only a physical curiosity but a programmable scaffold that merges optical actuation, directional transport control, and biofunctional surface engineering—opening avenues toward light-driven theranostic micro-objects, active diagnostic probes, and adaptive nanomedicine systems.

For transverse applications—particularly in microfluidics, biosensing, or biomedical environments—the particle size and long-term structural stability must be carefully considered. The current micrometric dimensions (~40 μm) are well suited for deterministic manipulation in laminar microflows, optical trapping, and shear-controlled transport, while avoiding rapid Brownian randomization. However, for in vivo or capillary-scale applications, size downscaling strategies (e.g., emulsification parameter optimization or microfluidic synthesis) could be implemented to access the sub-10 μm regime while preserving patch compartmentalization. Regarding stability, the azopolymer patches are covalently integrated within the composite architecture, reducing the risk of spontaneous detachment. Nevertheless, prolonged shear exposure, repeated photo-actuation cycles, or solvent variations may promote mechanical fatigue or interfacial weakening. Surface crosslinking of the azopolymer domains, interfacial compatibilization between PMMA and azopolymer phases, or thin protective coatings (e.g., silica or PEG layers) could enhance mechanical robustness and colloidal stability while preserving photo-responsiveness. Additionally, zeta-potential tuning or steric stabilization would mitigate aggregation under physiological ionic strengths. Addressing these size and durability considerations with biointerfacial engineering will therefore be essential to transition these light-programmable porcupine particles from proof-of-concept morphologies toward robust, application-ready active micro-objects for microfluidic, diagnostic, or theranostic systems.

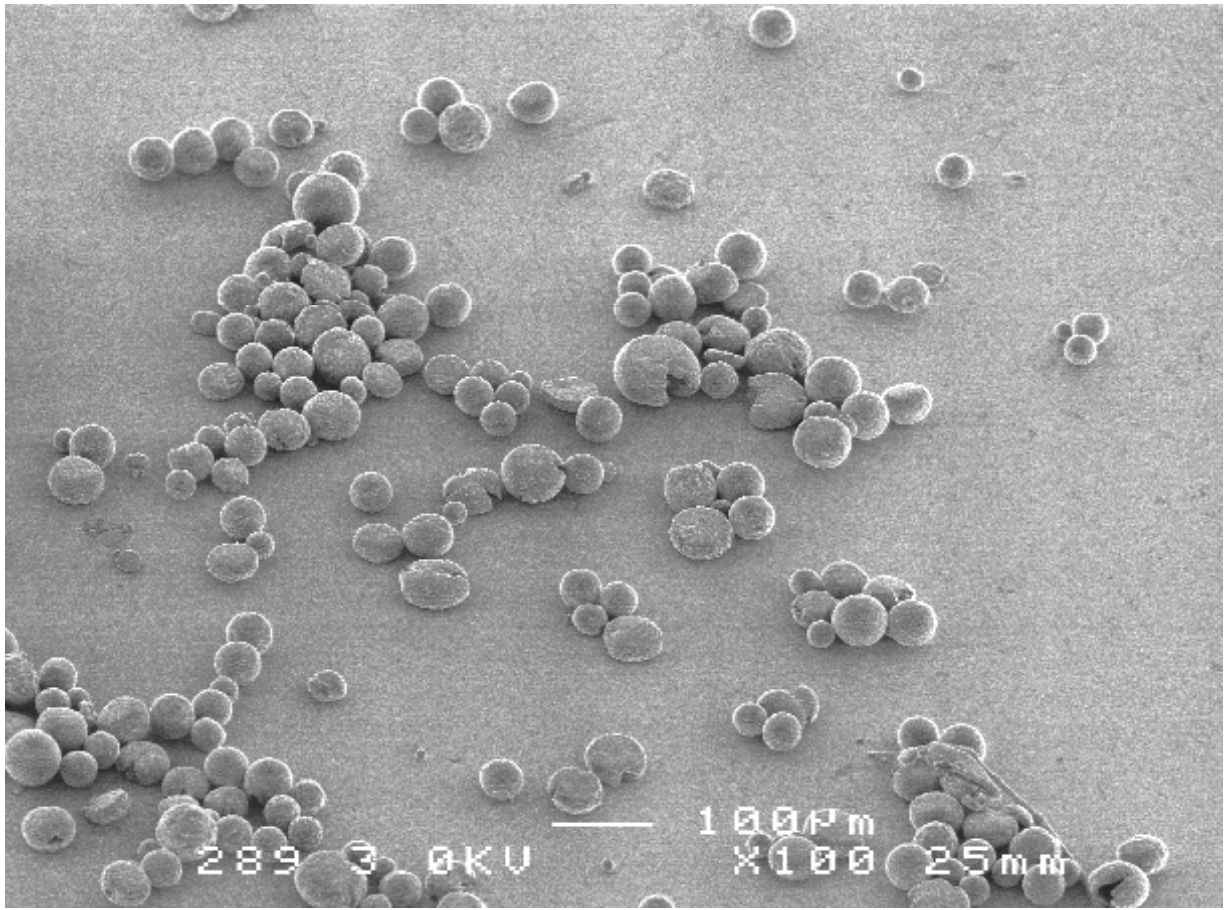
SI1: dispersion of PMMA particles coated with azopolymer patches on a glass substrate.

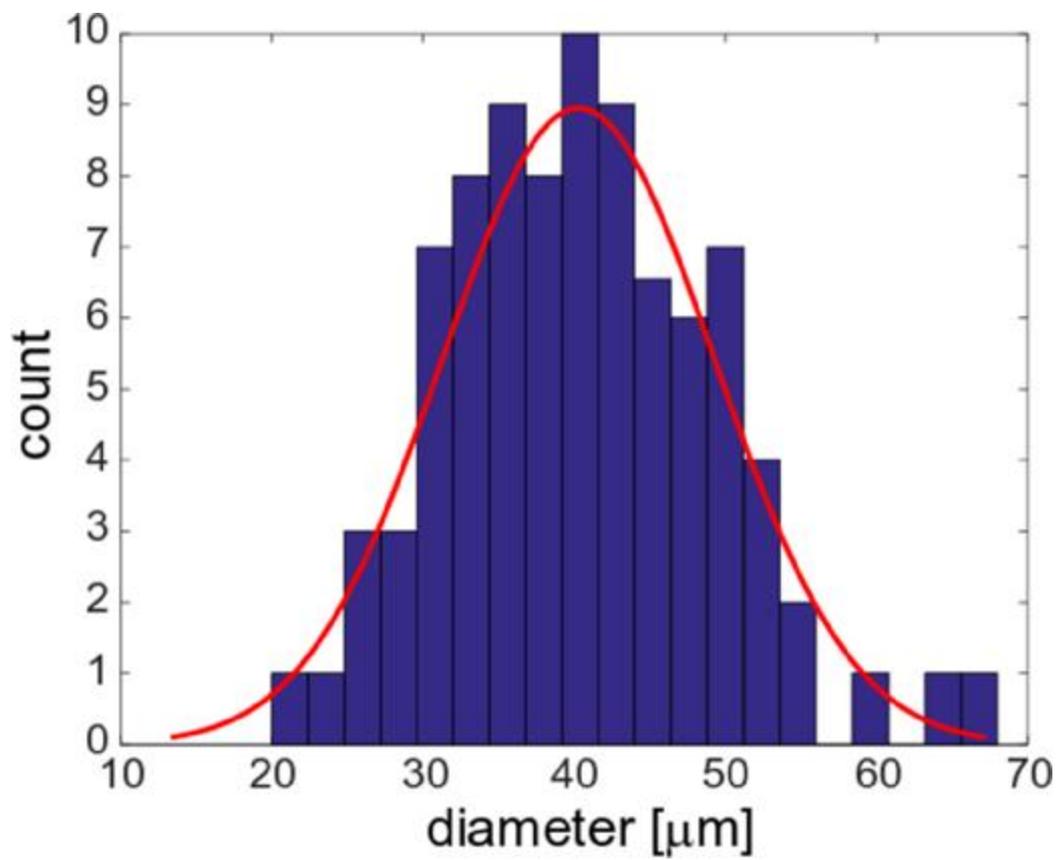

SI2: histogram with a Gaussian fit of the particle sizes (m= 40 ± 1 nm, s = 9 nm).

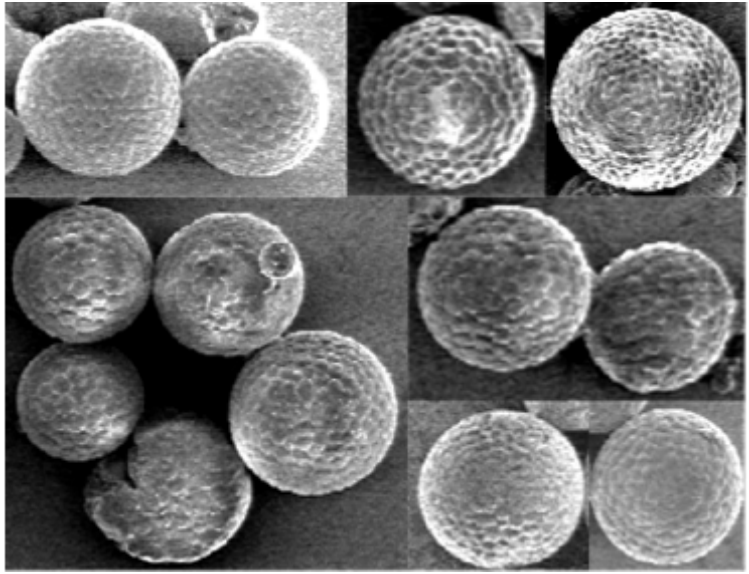

SI3: Details of the patchy particles containing azopolymer pacthes on the surface of the PMMA particles.

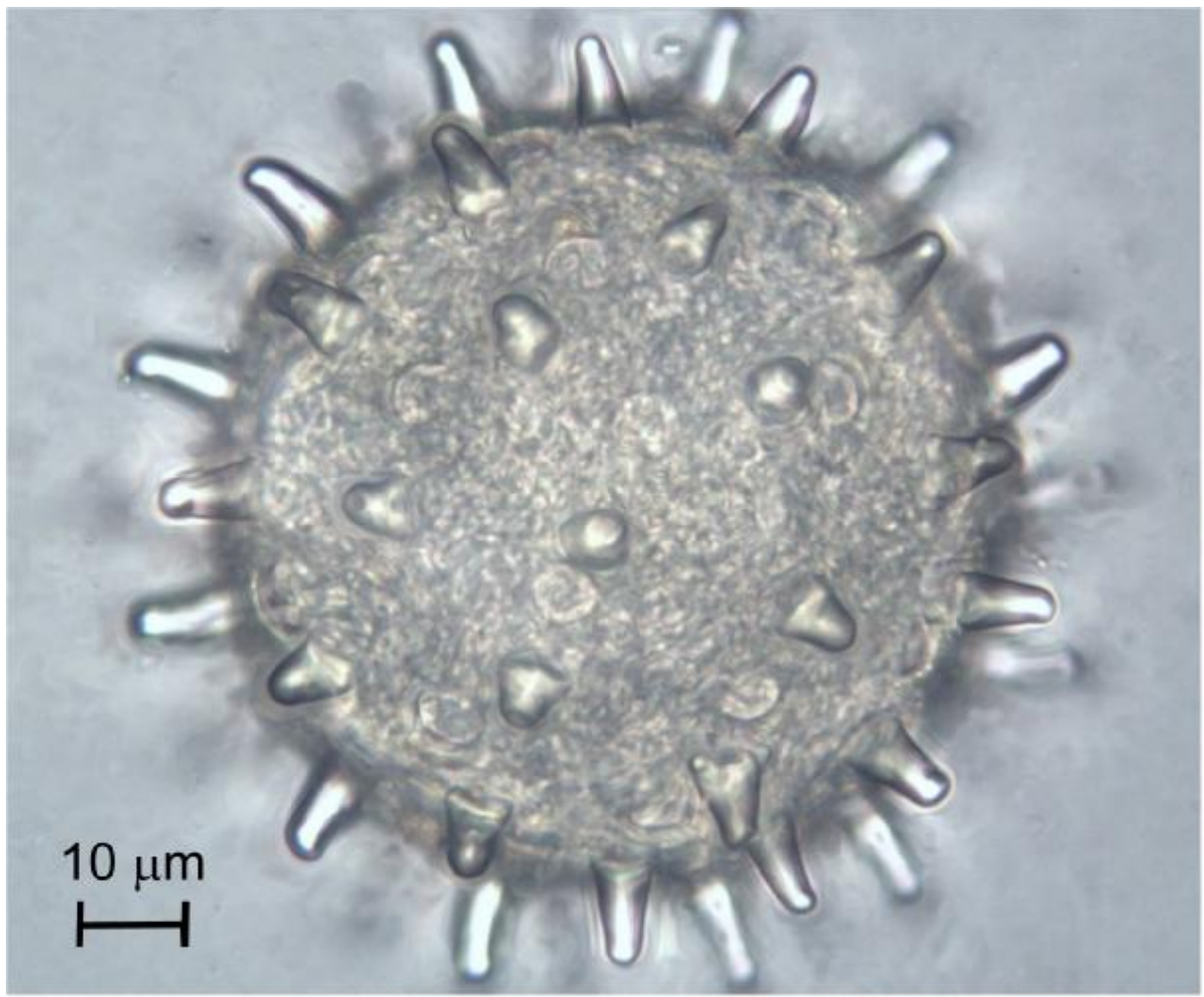

Fig. SI4: Image of a pollen grain

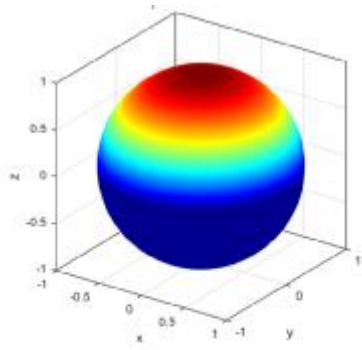
t = 0

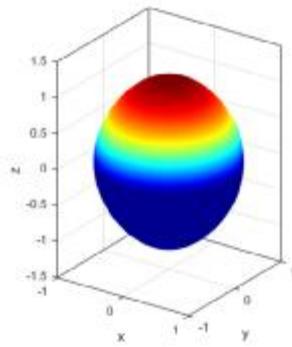
t = 0.25

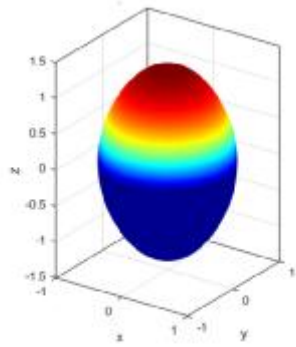
t = 0.5

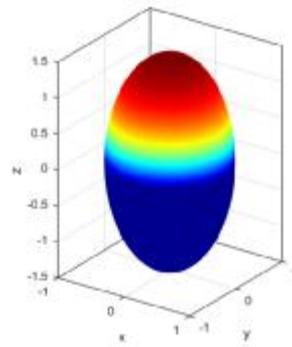
t = 1

**SI4: Stretching evolution of the particle under a linear polarization**

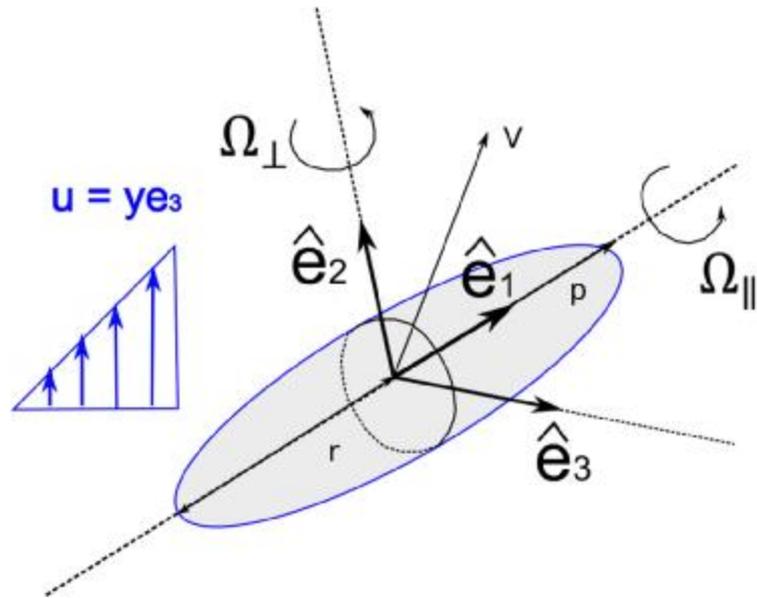

Fig. SI5: notation used in the calculation. r is the distance from center to pole of along the sphere or spheroid symmetry axis. The swimmer has self-generated translational and rotational velocities. $\Omega = \Omega_\parallel + \Omega_\perp$ respectively. Both particle interact with a background shear flow $u = ye_3$.

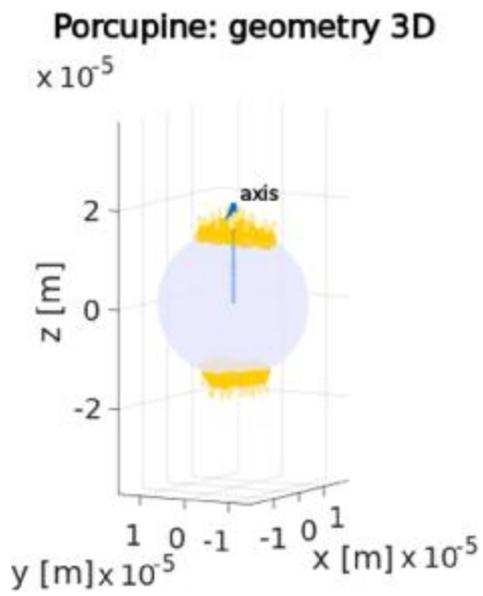 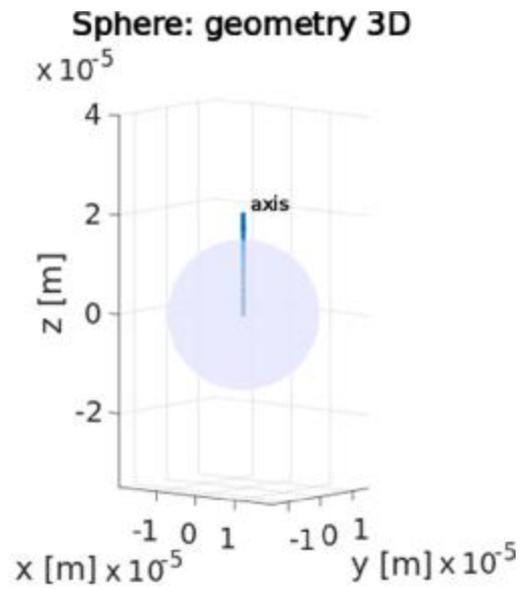
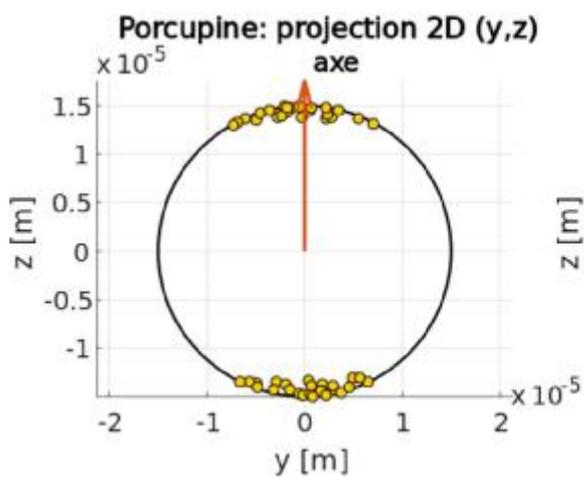 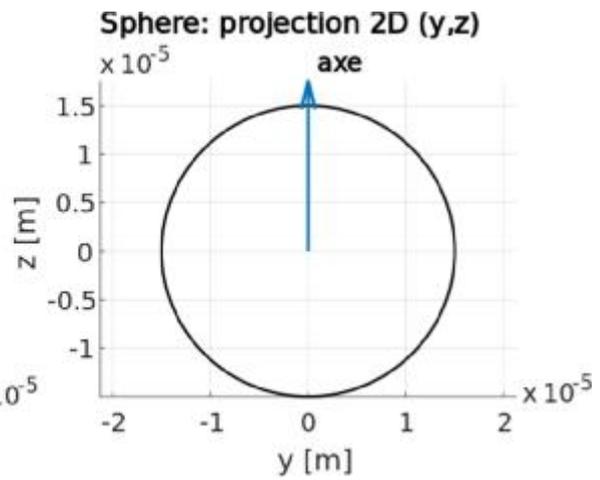

SI6: Geometry of the two particles used in the model for describing the behavior in a fluid with and without shear flow.

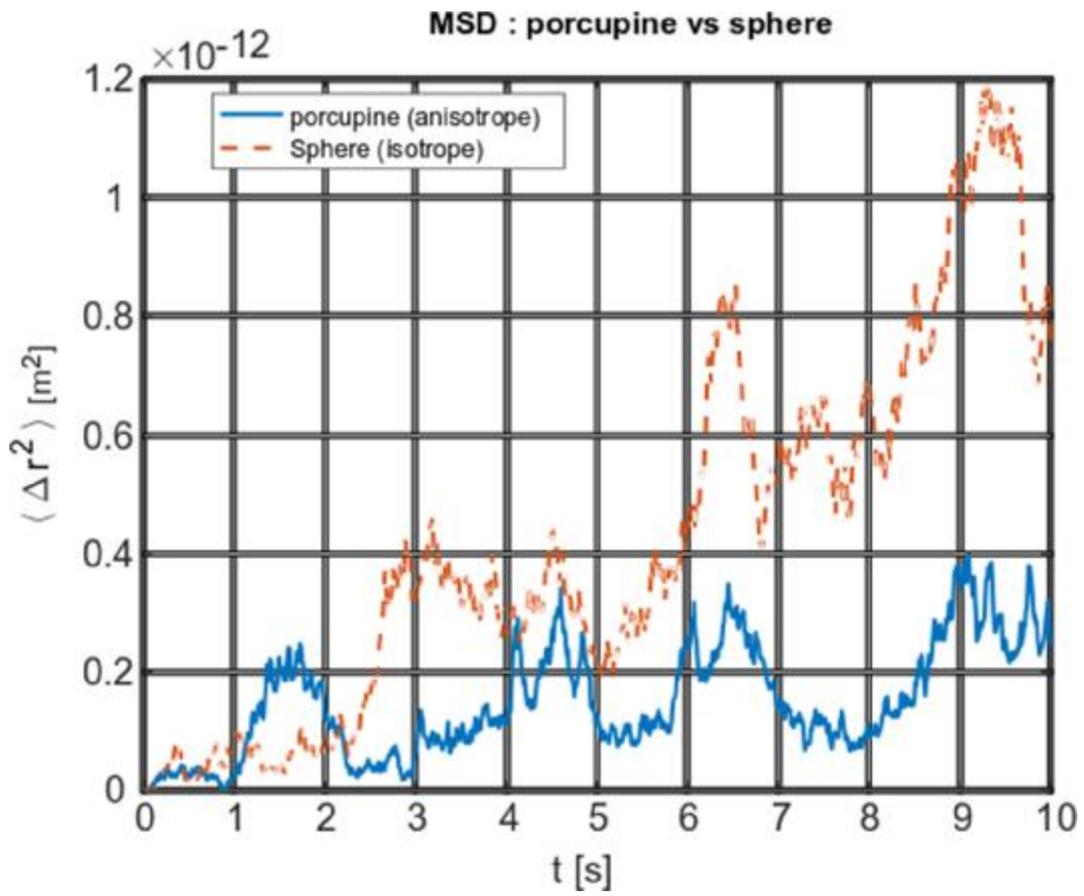

SI7: Comparison of the particle behavior in a fluid (water) without shear flow